\documentclass[11pt]{article}
\pdfoutput=1
\usepackage{jheppub}
\usepackage{bbm}
\usepackage{tikz}
\usetikzlibrary{calc}
\usepackage{graphicx,subcaption}
\usepackage{mathtools}
\usepackage{bm}
\usepackage{enumitem}
\usepackage{verbatim}
\usepackage{amssymb}
\usepackage{slashed}

\newcommand{\bfmtb}{\bfmtb}
\renewcommand{\bfmtb}{\lambda}

\DeclareMathOperator{\Tr}{Tr}

\DeclareMathAlphabet{\mathbfsf}{OT1}{cmss}{bx}{n}

\newcommand{\N}{\mathbb{N}}
\newcommand{\Z}{\mathbb{Z}}
\newcommand{\C}{\mathbb{C}}
\newcommand{\R}{\mathbb{R}}

\newcommand\be{\begin{equation}}
\newcommand\ee{\end{equation}}
\newcommand\beq{\begin{equation}}
\newcommand\eeq{\end{equation}}
\newcommand\bea{\begin{eqnarray}}
\newcommand\eea{\end{eqnarray}}

\renewcommand{\a}{\alpha}
\renewcommand{\b}{\beta}

\renewcommand{\t}{\tau}

\renewcommand{\tilde}{\widetilde}

\begin{document}

\title{Modular transform of free fermion generalised Gibbs ensembles and generalised power partitions}

\author{Max Downing}
\emailAdd{max.downing@kcl.ac.uk} 
\affiliation{Department of Mathematics, King's College London,\\
The Strand, London WC2R 2LS, U.K.}

\abstract{ 
In \cite{Downing:2021mfw} a conjecture for the modular transformation of the free fermion generalised Gibbs ensemble (GGE) was given where only the KdV charge associated to the weight four quasi primary field was inserted. In this paper we first generalise this conjecture to the case with an arbitrary, finite collection of KdV charges in the GGE. These GGEs are generalisations of the generating function of power partitions. We prove the conjectured transformation for the case with a finite number of charges inserted and discuss the case with an infinite number of charges.
}

\maketitle \flushbottom

\section{Summary of results}\label{sec:intro}

In \cite{hrj:8932} the modular transformation properties of the generalised eta function $\eta_s(\t)$, defined by
\be\label{eq:Zagiereta}
    \eta_s(\t) = e^{-\pi i \zeta(-s)\t}\prod_{m=1}^\infty \left(1 - e^{2\pi i m^s \t}\right)\;,
\ee
where $\zeta(z)$ is the Riemann zeta function, were studied. It was proved that under the modular S transformation, $\t\rightarrow-1/\t$, $\eta_s$ satisfies
\be
    \eta_s(\t) = (2\pi)^{\frac{s-1}{2}}\sqrt{i/\t} \prod_{\substack{\text{Im}(z)>0\\ \t z^s = \pm 1}} \eta_{1/s}(z)\;.
\ee
In the above expression the product is taken over the roots of the two polynomials $\t z^s=\pm 1$ such that the imaginary part of the roots are positive.

In this paper we study the modular transformation properties of a further generalisation of the Dedekind eta function where we replace the powers $m^s$ in the product in \eqref{eq:Zagiereta} with polynomials. For each $N=2,3,\dots$ we have a generalised eta function $\eta_{\a,\b}(\t,\a_1,\dots,\a_{2N-1})$ defined by
\be\label{eq:geneta}\begin{split}
    \eta_{\a,\b}(\t,\a_1,\dots,\a_{2N-1}) =& \exp\left(-\pi i \sum_{n=1}^N \a_{2n-1} \t^{2n-1} \zeta(1-2n,1-\a) \right) \\
    &\times \prod_{\substack{k \in \Z+\a\\ k>0}}^\infty \left( 1 - e^{2\pi i (\b + \sum_{n=1}^N \a_{2n-1} (\t k)^{2n-1})} \right)\;,
\end{split}\ee
where $\zeta(s,a)$ is the Hurwitz zeta function, defined in \eqref{eq:Hurwitzzeta}. The product in \eqref{eq:geneta} converges for Im$(\a_{2N-1}\t^{2N-1})>0$ and the Hurwitz zeta function is defined for $a \neq 0,-1,\dots$. While $\eta_{\a,\b}$ can be defined for $\a \in \R\setminus\{1,2,\dots\}$ and $\b \in \R$, we will restrict $(\a,\b)$ to the four pairs $(0,0)$, $\left(0,\tfrac{1}{2}\right), \left(\tfrac{1}{2},0\right)$ and $\left(\tfrac{1}{2},\tfrac{1}{2}\right)$ since these functions are related to the Generalised Gibbs ensemble of a free fermion in 2 dimensions as described below. We will additionally assume throughout that Im$(\t)>0$ but place no restriction on the $\a_{2n-1}\t^{2n-1}$ for $n<N$.

This is not the most general order $2N-1$ polynomial that one could include in \eqref{eq:geneta}, we are restricting to odd polynomials. This restriction is because we will be interested in the link between these generalised eta functions and the generalised Gibbs ensembles studied in \cite{Downing:2021mfw} (we will say more about this link below). The only polynomials that occur in these generalised Gibbs ensembles are odd and in addition have a non-vanishing linear term i.e. $\a_1=1$. Note that we can still relate our results to those of \cite{hrj:8932} by setting $\a_{2n-1} = 0$ for $n<N$, $\a_{2N-1} = 1$ and $\t^{2N-1} = \t'$, the condition Im$(\a_{2N-1}\t^{2N-1}) > 0$ implies Im$(\t') > 0$.
 
In this paper we will prove the following transformation properties for the generalised eta function \eqref{eq:geneta}
\be\label{eq:transform1}\begin{split}
    &\eta_{\a,\b}(\t,\a_1,\dots,\a_{2N-1}) = \\
    &\sqrt{2}^{\delta_{\a,\frac{1}{2}}\delta_{\b,0} - \delta_{\a,0}\delta_{\b,\frac{1}{2}}} \sqrt{i/\t}^{\delta_{\a,0}\delta_{\b,0}}\exp\left( \int_0^\infty \log(1 - e^{2\pi i\b}e^{2\pi i(\sum_{n=1}^N \a_{2n-1} (\t x)^{2n-1})}) dx \right)\\
    &\times \prod_{n\in\Z+\b} \prod_{\substack{z_j(n) \\ \text{Im}(z_j(n))>0}} (1-e^{2\pi i (\a + z_j(n))})\;,
\end{split}\ee
where the $z_j(x)$, $j=1,\dots,2N-1$ are the roots of the polynomial
\be
    \sum_{n=1}^N \a_{2n-1} (\t z)^{2n-1} = x\;.
\ee
This transformation formula is valid for the domain given by Im$(\t)>0$, $\a_1\neq0$ and Im$(\a_{2N-1}\t^{2N-1})$. For the special case $\a_{2n-1} = 0$ for all $n<M\leq N$ and $\a_{2M-1} \neq 0$ the transformation can be found by starting with \eqref{eq:transform1} with $\a_1\neq0$ and taking the limit $\a_1\rightarrow0$. Note that if one naively sets $\a_{2n-1} = 0$ for all $n<M\leq N$ and $\a_{2M-1} \neq 0$ in \eqref{eq:transform1} then the transform is incorrect as factors corresponding to the roots $z_j(0)$ will be missed.

We are interested in the transformation properties of $\eta_{\a,\b}$ because of its relation to the generalised Gibbs ensembles (GGEs) of free fermions. In \cite{Downing:2021mfw} the modular properties of these GGEs for free fermions was studied. These GGEs were denoted in \cite{Downing:2021mfw} by
\be\label{eq:GGE}
    \Tr_{\text{NS/R},\pm} \left( e^{2\pi i(\t I_1 + \sum_{n=2} \a_{2n-1} \t^{2n-1} I_{2n-1})} \right)\;,
\ee
where the conserved charges in the GGE are the KdV charges, $I_{2n-1}$. A definition for the traces $\Tr_{\text{NS/R},\pm}$ can be found in appendix A of \cite{Downing:2021mfw}. These GGEs are related to the generalised eta function as follows
\be\label{eq:etaGGE}\begin{split}
    &\eta_{0,0}(\t,1,\a_3,\dots,\a_{2N-1}) = \Tr_{\text{R},-}\left( e^{2\pi i(\t I_1 + \sum_{n=2}^N \t^{2n-1}\a_{2n-1} I_{2n-1})} \right)\;,\\
    &\eta_{0,1/2}(\t,1,\a_3,\dots,\a_{2N-1}) = \Tr_{\text{R},+}\left( e^{2\pi i(\t I_1 + \sum_{n=2}^N \t^{2n-1}\a_{2n-1} I_{2n-1})} \right)\;,\\
    &\eta_{1/2,0}(\t,1,\a_3,\dots,\a_{2N-1}) = \Tr_{\text{NS},-}\left( e^{2\pi i(\t I_1 + \sum_{n=2}^N \t^{2n-1}\a_{2n-1} I_{2n-1})} \right)\;,\\
    &\eta_{1/2,1/2}(\t,1,\a_3,\dots,\a_{2N-1}) = \Tr_{\text{NS},+}\left( e^{2\pi i(\t I_1 + \sum_{n=2}^N \t^{2n-1}\a_{2n-1} I_{2n-1})} \right)\;.
\end{split}\ee
In \cite{Downing:2021mfw} we looked at the case with just the $I_3$ charge inserted (only $\a_1$ and $\a_3$ non zero) and using the thermodynamic Bethe ansatz (TBA) we conjectured the modular transformation for \eqref{eq:GGE} under $\t \rightarrow -1/\t$. Our conjecture was the transformation \eqref{eq:transform1} for $\eta_{\a,\b}(\t,1,\a_3)$. Note that in \cite{Downing:2021mfw} we only studied the three cases $(\a,\b) = \left(0,\frac{1}{2}\right), \left(\frac{1}{2},0\right), \left(\frac{1}{2},\frac{1}{2}\right)$ but the conjecture can naturally be extended to include $(\a,\b)=(0,0)$. In the companion paper \cite{DowningWatts} we show how the transformations \eqref{eq:transform1} can again be derived from the TBA. In \cite{DowningWatts} we also provide a physical interpretation for the transformed GGEs in terms of a line defect.

In this paper we give a mathematical proof for the transformation \eqref{eq:transform1}. We will use the techniques from \cite{hrj:8932} in our proof. By writing these expressions as sums over 1 dimensional lattices we can use Poisson summation to find the modular transform. This approach involves computing a Fourier transform and then deforming the integration contour to pick up poles, much in the same way one deforms the integration contour in the TBA equations to pick up the excited states. This suggests that it may be possible to adapt this method to study the GGEs of other 2d CFTs, such as the minimal models. However the eigenvalues of the KdV charges are not known in general for other theories so explicit calculations cannot be done at this stage. 

The paper is organised as follows. The main details of the proof of the transformation \eqref{eq:transform1} are presented in section \ref{sec:proof}. We then briefly discuss how this proof can be applied to the case with an infinite collection of non zero $\a_{2n-1}$ in the GGE \eqref{eq:GGE} in section \ref{sec:infinite}. Two appendices with technical results needed for the proof in section \ref{sec:proof} have also been included.

\section{Proof of modular transform for finite charges}\label{sec:proof}

Here we will prove the modular transform given in \eqref{eq:transform1}. To do this we will first restrict to the case where $\a_1\neq 0$ and Im$(\a_{2n-1}\t^{2n-1}) \geq0$ for $n<N$ and prove the transform $\eqref{eq:transform1}$. We can then use analytic continuation to extend the transform \eqref{eq:transform1} to the larger domain given by Im$(\t)>0$ and Im$(\a_{2N-1}\t^{2N-1})>0$. To begin the proof take the logarithm of $\eta_{\a,\b}$ so the product becomes a sum. If we introduce a suitable parameter, $\lambda$ in \eqref{eq:lambda}, that we differentiate with respect to twice we will be able to write this as a sum over the full lattice $\Z+\a$. This allows us to use Poisson summation to find the transformation. After computing the required Fourier transform for the Poisson summation we can then integrate twice with respect to $\lambda$. Fixing the integration constants using asymptotic expansions in $\lambda$ will give us the final form for the transformation.

We start by repeating the definition of the generalisation of the Dedekind eta function given in \eqref{eq:geneta}
\be\begin{split}
    \eta_{\a,\b}(\t,\a_1,\dots,\a_{2N-1}) =& \exp\left(-\pi i \sum_{n=1}^N \a_{2n-1} \t^{2n-1} \zeta(1-2n,1-\a)\right) \\
    &\times \prod_{\substack{k \in \Z+\a\\ k>0}}^\infty \left( 1 - e^{2\pi i (\b + \sum_{n=1}^N \a_{2n-1} (\t k)^{2n-1})} \right)\;,
\end{split}\ee
As discussed below \eqref{eq:geneta} we will look at the four $(\a,\b)$ pairings $(0,0), \left(0,\frac{1}{2}\right), \left(\frac{1}{2},0\right)$ and $\left(\frac{1}{2},\frac{1}{2}\right)$ and the $\t$ and $\a_{2N-1}$ variables will be restricted to the domains Im$(\t)>0$ and Im$(\t^{2N-1}\a_{2N-1}) > 0$. Here we will also impose $\a_1\neq 0$ and Im$(\t^{2N-1}\a_{2N-1}) \geq 0$ for $n<N$ and analytically continue to the larger domain at the end. 

In order to prove the transformation \eqref{eq:transform1} we will take the logarithm of $\eta_{\a,\b}$. This will turn the product into a sum that we can then perform Poisson summation on. Define the function $H_{\a,\b}$ as the logarithm of $\eta_{\a,\b}$
\be\label{eq:H}\begin{split}
    H_{\a,\b}(\t,\a_1,\dots,\a_{2N-1}) =& -\pi i \sum_{n=2}^N \a_{2n-1}\t^{2n-1}\zeta(1-2n,1-\a)\\ 
    &+\sum_{\substack{k\in\Z+\a \\ k>0}} \log\left( 1 - e^{2\pi i (\b + \sum_{n=1}^N \a_{2n-1} (\t k)^{2n-1})} \right)\;.
\end{split}\ee
We will pick the branch cut of the logarithm such that $\log(-1)=-\pi i$. However since we will exponentiate the final result the choice of branch doesn't matter. We want to use Poisson summation on the series in $H_{\a,\b}$. However the series is only over the half lattice $k \in \Z+\a, k>0$, we need to extend it to the whole lattice ($k \in \Z+\a$) in order to use the summation formula. To do this first note that
\be\label{eq:integersum}\begin{split}
    &2\sum_{\substack{k \in \Z+\a \\ k>0}} \log\left( 1 - e^{2\pi i (\b + \sum_{n=1}^N \a_{2n-1} (\t k)^{2n-1})} \right)\\
    &= \sum_{\substack{k \in \Z+\a \\ k>0}} \log\left( 1 - e^{2\pi i (\b + \sum_{n=1}^N \a_{2n-1} (\t k)^{2n-1})} \right) + \sum_{\substack{k \in \Z+\a \\ k<0}} \log\left( 1 - e^{-2\pi i (\b + \sum_{n=1}^N \a_{2n-1} (\t k)^{2n-1})} \right)\\
    &= \sum_{\substack{k \in \Z+\a \\ k>0}} \log\left( 1 - e^{2\pi i (\b + \sum_{n=1}^N \a_{2n-1} (\t k)^{2n-1})} \right) \\
    &+ \sum_{\substack{k \in \Z+\a \\ k<0}} \left( -2\pi i \left(\frac{1}{2} + \b + \sum_{n=1}^N \a_{2n-1} (\t k)^{2n-1}\right) + \log\left( 1 - e^{2\pi i (\b + \sum_{n=1}^N \a_{2n-1} (\t k)^{2n-1})} \right)\right)\;.
\end{split}\ee
Since we are only considering $(\a,\b) = (0,0), \left(0,\frac{1}{2}\right), \left(\frac{1}{2},0\right), \left(\frac{1}{2},\frac{1}{2}\right)$ we can swap $-\a$ for $\a$ and $-\b$ for $\b$ in the above series. Note that in the final line we have used the fact that we chose a branch cut for the logarithm such that $\log(-1) = -\pi i$. The final sum in \eqref{eq:integersum} is convergent but if we split the series into two separate pieces
\be\label{eq:divergentsums}
    \sum_{\substack{k \in \Z+\a \\ k>0}}\left(\frac{1}{2} + \b + \sum_{n=1}^N \a_{2n-1} (\t k)^{2n-1}\right) \;,\quad \sum_{\substack{k \in \Z+\a \\ k>0}}\log\left( 1 - e^{2\pi i (\b + \sum_{n=1}^N \a_{2n-1} (\t k)^{2n-1})} \right)\;,
\ee
each series is individually divergent. However if we formally take the series 
\be
\sum_{\substack{k \in \Z+\a \\ k>0}}\log\left( 1 - e^{2\pi i (\b + \sum_{n=1}^N \a_{2n-1} (\t k)^{2n-1})} \right)\;,
\ee
and the convergent series 
\be
\sum_{\substack{k \in \Z+\a \\ k>0}} \log\left( 1 - e^{2\pi i (\b + \sum_{n=1}^N \a_{2n-1} (\t k)^{2n-1})} \right)\;,
\ee
and sum them we get a series over the whole lattice which we can then apply Poisson summation to, at least formally. (Note that when $\a=0$ we have a $k=0$ term which we will have to deal with separately.) We can get this series over the full lattice by effectively differentiating twice with respect to $\sum_{n=1}^N \a_{2n-1} (\t k)^{2n-1}$ to remove this divergent series in the last line of \eqref{eq:integersum}. We do this by introducing a positive real parameter $\lambda$ into the sum in \eqref{eq:H} as follows
\be\label{eq:lambda}
    \sum_{k \in \Z+\a,k>0} \log\left( 1 - e^{2\pi i\b}e^{2\pi i \lambda \sum_{n=1}^N \a_{2n-1} (\t k)^{2n-1}} \right)\;.
\ee
The series converges exponentially so we can differentiate term wise with respect to $\lambda$. Differentiating twice with respect to $\lambda$ yields
\be\label{eq:latticesumlambda}\begin{split}
    &2\sum_{\substack{k \in \Z+\a \\ k>0}}^\infty \frac{\partial^2}{\partial \lambda^2} \log\left( 1 - e^{2\pi i\b}e^{2\pi i \lambda \sum_{n=1}^N \a_{2n-1} (\t k)^{2n-1}} \right) \\
    &= \sum_{k \in \Z+\a} \frac{\partial^2}{\partial \lambda^2} \log\left( 1 - e^{2\pi i\b}e^{2\pi i \lambda \sum_{n=1}^N \a_{2n-1} (\t k)^{2n-1}} \right) +\delta_{\a,0}\delta_{\b,0}\frac{1}{\lambda^2}\;.
\end{split}\ee
The second sum can be written over the whole lattice because the terms in the sum decay exponentially as $k\rightarrow \pm\infty$. Note that when $(\a,\b)=\left(0,\frac{1}{2}\right)$ the term $k=0$ is independent of $\lambda$ and so vanishes. When $(\a,\b)=(0,0)$ the $k=0$ term can be defined by taking the limit $k\rightarrow0$ to get a finite value. However since our original series is over $k>0$ we have to remove the $k=0$ term from the lattice sum which we have done in \eqref{eq:latticesumlambda}. Hence the sum of the series over $k>0$ and $k<0$ can be written as a sum over $k\in\Z+\a$. We now use the Poisson summation formula on this series.

The Poisson summation formula on the shifted integer lattice is
\be\label{eq:poissonsum}
    \sum_{\nu \in \Z+\a} f(\nu t) = \frac{1}{t} \sum_{m \in \Z} \tilde f(m/t)e^{-2\pi im\a}\;,
\ee
where
\be
    \tilde f(y) = \int_{-\infty}^\infty f(t)e^{2\pi it y}dt\;.
\ee
In order to use the summation formula on the series
\be\label{eq:fulllattice}
    \sum_{k \in \Z+\a} \frac{\partial^2}{\partial \lambda^2} \log\left( 1 - e^{2\pi i\b}e^{2\pi i \lambda \sum_{n=1}^N \a_{2n-1} (\t k)^{2n-1}} \right)\;,
\ee
we first define the function
\be
    f_\lambda(t) = \log\left( 1 - e^{2\pi i\b}e^{2\pi i\lambda \sum_{n=1}^N \a_{2n-1} (it)^{2n-1}} \right)\;,
\ee
The series
\be\label{eq:fderivlatticesum}
    \sum_{\nu \in \Z+\a} \frac{\partial^2}{\partial \lambda^2}f_\lambda(\nu t)\;,
\ee
converges for Im$(\a_{2N-1}(it)^{2N-1})>0$ and is related to \eqref{eq:fulllattice} by setting $t = -i\t$. We will use Poisson summation on \eqref{eq:fderivlatticesum} for $t>0$ and then analytically continue the result into the complex plane. We need to compute the Fourier transform of $F_\lambda(t) = \frac{\partial^2}{\partial \lambda^2}f_\lambda(t)$ with respect to $t$. First we define
\be\label{eq:pdef}
    p_\lambda(t) = 1 - e^{2\pi i\b}e^{2\pi i \lambda \sum_{n=1}^N \a_{2n-1} (it)^{2n-1}}\;,
\ee
so $f_\lambda(t) = \log(p_\lambda(t))$ and the Fourier transform of $F_\lambda(t)$ is
\be
    \tilde F_\lambda(y) = \int_{-\infty}^\infty \frac{\partial^2}{\partial \lambda^2}\log(p_\lambda(t)) e^{2\pi it y}dt\;.
\ee
For $y \neq 0$ we integrate by parts and drop the boundary term which vanishes
\be
    \tilde F_\lambda(y) = \int_{-\infty}^\infty \frac{\partial^2}{\partial \lambda^2} \log(p_\lambda(t)) e^{2\pi it y}dt = -\frac{1}{2\pi i y}\int_{-\infty}^\infty \frac{\partial^2}{\partial \lambda^2} \left(\frac{p_\lambda'(t)}{p_\lambda(t)}\right) e^{2\pi ity}dt 
\ee
where $p_\lambda'(t) = \frac{\partial}{\partial t}p_\lambda(t)$. In order to compute the integral 
\be\label{eq:pderivoverp}
    -\frac{1}{2\pi i y}\int_{-\infty}^\infty \frac{\partial^2}{\partial \lambda^2} \left(\frac{p_\lambda'(t)}{p_\lambda(t)}\right) e^{2\pi ity}dt\;,
\ee
we close the contour either in the upper or lower half plane, depending on the sign of $y$, and pick up the residues. These residues come from the zeros of $p(t)$. (For a generic function $p(t)$ the integral \eqref{eq:pderivoverp} will also have contributions from the poles of $p(t)$, but $p(t)$ as defined in \eqref{eq:pdef} has no poles.) 

For $y>0$ we close the contour in the upper half plane. If $p(t)$ has a zero of order $M$ at $t = t_0$, which will depend on $\lambda$, in the upper half plane then the contribution to the integral is 
\be
    -\frac{M}{y} \frac{\partial^2}{\partial \lambda^2} e^{2\pi i t_0 y}\;.
\ee
This is because if we integrate anticlockwise around a contour that encloses the zero $t_0$ and no other zeros we can move the $\lambda$ derivatives outside the integral and evaluate it to find
\be
    -\frac{1}{2\pi i y} \frac{\partial^2}{\partial \lambda^2} \int_{C(t_0)} \frac{p_\lambda'(t)}{p_\lambda(t)} e^{2\pi ity}dt = -\frac{M}{y} \frac{\partial^2}{\partial \lambda^2} e^{2\pi i t_0 y}\;.
\ee
(Here $C(t_0)$ is the anticlockwise contour described above.) The zero $t_0$ must satisfy the following polynomial
\be
    \sum_{n=1}^N \a_{2n-1} (it)^{2n-1} = \frac{n}{\lambda}\;,\quad n \in \Z+\b \;,
\ee
hence the order of the zero, $M$, is just the degeneracy of the root $t_0$. Hence the Fourier transform is given by the sum over all zeros of $p_\lambda(t)$ in the upper half plane
\be\label{eq:positivey}
    \tilde F_\lambda(y) = -\sum_{n\in\Z+\b}\sum_{\substack{\tilde z_j(n/\lambda) \\ \text{Im}(\tilde z_j(n/\lambda))>0}} \frac{1}{y} \frac{\partial^2}{\partial \lambda^2} e^{2\pi i \tilde z_j(n/\lambda)y}\;,
\ee
where $\tilde z_j(x)$, $j=1,\dots,2N-1$, are the roots of
\be\label{eq:xpoly}
    \sum_{n=1}^N \a_{2n-1} (i\tilde z)^{2n-1} = x\;,
\ee
If we have a degenerate root then we sum over each copy separately in \eqref{eq:positivey}. Similarly for $y<0$ we close the contour in the lower half plane and the Fourier transform is
\be\label{eq:negativey}
    \tilde F_\lambda(y) = \sum_{n\in\Z+\b}\sum_{\substack{\tilde z_j(n\lambda) \\ \text{Im}(\tilde z_j(n/\lambda))<0}} \frac{1}{y} \frac{\partial^2}{\partial \lambda^2} e^{2\pi i \tilde z_j(n/\lambda)y}\;,
\ee
where again $\tilde z_j(n/\lambda)$, $j=1,\dots,2N-1$, are the roots of \eqref{eq:xpoly}. Note that here we are now integrating around a clockwise contour so we pick up a minus sign. For $y=0$ the Fourier transform is simply given by the integral
\be\label{eq:zeroy}
    \tilde F_\lambda(0) = \int_{-\infty}^\infty \frac{\partial^2}{\partial \lambda^2} \log(p_\lambda(u)) du = \int_{-\infty}^\infty \frac{\partial^2}{\partial \lambda^2} \log\left(1 - e^{2\pi i\b}e^{2\pi i \lambda \sum_{n=2}^N \a_{2n-1} (iu)^{2n-1}} \right) du\;.
\ee
If we split the right hand side of \eqref{eq:poissonsum} into a sum over $m>0$, a sum over $m<0$ and the $m=0$ term and use the expressions \eqref{eq:positivey}, \eqref{eq:negativey} and \eqref{eq:zeroy} we have
\be\begin{split}
    &\sum_{k \in \Z+\a} \frac{\partial^2}{\partial \lambda^2} \log\left( 1 - e^{2\pi i\b}e^{2\pi i \lambda \sum_{n=1}^N \a_{2n-1} (\t k)^{2n-1}} \right)\\
    &= \frac{1}{t}\int_{-\infty}^\infty \frac{\partial^2}{\partial \lambda^2} \log\left(1 - e^{2\pi i\b}e^{2\pi i\lambda \sum_{n=1}^N \a_{2n-1} (iu)^{2n-1}} \right) du \\
    &- \sum_{m=1}^\infty \sum_{n\in\Z+\b} \left(\sum_{\substack{\tilde z_j(n/\lambda) \\ \text{Im}(\tilde z_j(n/\lambda))>0}} \frac{\partial^2}{\partial \lambda^2} \frac{e^{2\pi i m (\a + \tilde z_j(n/\lambda)/t)}}{m} + \sum_{\substack{\tilde z_j(n/\lambda) \\ \text{Im}(\tilde z_j(n/\lambda))<0}} \frac{\partial^2}{\partial \lambda^2} \frac{e^{2\pi i m (\a - \tilde z_j(n/\lambda)/t)}}{m} \right)\;.
\end{split}\ee
Note we have written the sum over $m<0$ as a sum over $m>0$. We are free to swap $-\a$ for $\a$ when doing this since $\a = 0$ or $\a = \frac{1}{2}$. We can swap the $n$ and $m$ summations in the last line since the sums converge exponentially and then evaluate the sums over $m$ to get a logarithm
\be\label{eq:psumx}\begin{split}
    &\sum_{k \in \Z+\a} \frac{\partial^2}{\partial \lambda^2} \log\left( 1 - e^{2\pi i\b}e^{2\pi i \lambda \sum_{n=1}^N \a_{2n-1} (\t k)^{2n-1}} \right)\\
    &= \frac{1}{t}\int_{-\infty}^\infty \frac{\partial^2}{\partial \lambda^2} \log\left(1 - e^{2\pi i\b}e^{2\pi i\lambda \sum_{n=1}^N \a_{2n-1} (iu)^{2n-1})}\right) du \\
    &+\sum_{n\in\Z+\b} \left(\sum_{\substack{\tilde z_j(n/\lambda) \\ \text{Im}(\tilde z_j(n/\lambda))>0}} \frac{\partial^2}{\partial \lambda^2} \log\left(1-e^{2\pi i (\a + \tilde z_j(n/\lambda)/t)}\right) + \sum_{\substack{\tilde z_j(n) \\ \text{Im}(\tilde z_j(n))<0}} \frac{\partial^2}{\partial \lambda^2} \log\left(1-e^{2\pi i(\a - \tilde z_j(n)/t)}\right)) \right)\;.
\end{split}\ee
We will now rewrite the above expressions. We set $z_j(x) = i \tilde z_j(x)/\t$ where $\t = it$. Recall that $t>0$ so Im$(\t)>0$ and the sign of Im$(z_j)$ is the same as the sign of Im$(\tilde z_j)$. Hence \eqref{eq:psumx} becomes
\be\label{eq:psumz}\begin{split}
    &\sum_{\nu \in \Z+\a} \frac{\partial^2}{\partial \lambda^2} \log(1 - e^{2\pi i\b}e^{2\pi i \lambda \sum_{n=1}^N \a_{2n-1} (\nu\t)^{2n-1}})\\
    &= \frac{i}{\t}\int_{-\infty}^\infty \frac{\partial^2}{\partial \lambda^2} \log\left(1 - e^{2\pi i\b}e^{2\pi i\lambda \sum_{n=1}^N \a_{2n-1} (iu)^{2n-1}} \right) du \\
    &+\sum_{n\in\Z+\b} \left(\sum_{\substack{z_j(n/\lambda) \\ \text{Im}(z_j(n/\lambda))>0}} \frac{\partial^2}{\partial \lambda^2} \log \left(1-e^{2\pi i (\a + z_j(n/\lambda))} \right) + \sum_{\substack{z_j(n/\lambda) \\ \text{Im}(z_j(n/\lambda))<0}} \frac{\partial^2}{\partial \lambda^2} \log \left(1-e^{2\pi i(\a - z_j(n/\lambda))}\right) \right)\;,
\end{split}\ee
where the $z_j(x)$, $j=1,\dots,2N-1$, are the roots of
\be\label{eq:zpoly}
    \sum_{n=1}^N \a_{2n-1} (\t z)^{2n-1} = x\;.
\ee
As was shown in \eqref{eq:latticesumlambda} the first sum in \eqref{eq:psumz} can be written as a sum over the half lattice with an extra term from $\nu=0$ when $(\a,\b)=(0,0)$. On the half lattice the sum converges without the $\lambda$ differentiation so we can move the derivatives outside the sum. 

For the integral in \eqref{eq:psumz} we change the integration variable to $x = iu/\t$. Recall $\t = it$ and $t>0$ so the integral is still along the positive real axis. As a function of $x$ the integrand is an even function so the integral over the real line can be written as an integral over the positive reals. Now if we remove the $\lambda$ derivatives from the integrand the integral converges so we can move the derivatives outside of the integral.

Finally we look at the sum over the roots $z_j(x)$ in \eqref{eq:psumz}. As discussed in appendix \ref{sec:roots} one of the roots of \eqref{eq:zpoly} vanishes when $x=0$. We label this root $z_0(x)$ and furthermore in appendix \ref{sec:roots} we showed it is an odd function of its argument and Im$(z_0(x))<0$ for $x>0$. The other $2N-2$ roots come in pairs and in appendix \ref{sec:roots} are labeled $z_j^\pm(x), j = 1,\dots,N-1$. We labelled them such that Im$(z_i^+(x))>0$ and Im$(z_i^-(x))<0$ for $x>0$ and we have the relation $z_i^+(x) = - z_i^-(-x)$. Using these results for the roots allows us to rewrite the two sums on the bottom line of \eqref{eq:psumz} as
\be
    2\frac{\partial^2}{\partial \lambda^2}\sum_{\substack{n\in\Z+\b\\ n>0}} \left(\sum_{\substack{z_j(n/\lambda) \\ \text{Im}(z_j(n/\lambda))>0}} \log\left(1-e^{2\pi i (\a + z_j(n/\lambda))}\right) + \sum_{\substack{z_j(n/\lambda) \\ \text{Im}(z_j(n/\lambda))<0}} \log\left(1-e^{2\pi i(\a - z_j(n/\lambda))}\right) \right)\;,
\ee
where again we can move the $\lambda$ derivatives outside of the sum since the sums converge exponentially even without differentiating term wise. 

Putting all this together means \eqref{eq:psumz} becomes
\be\label{eq:psum}\begin{split}
    &\frac{\partial^2}{\partial \lambda^2} \sum_{\substack{\nu \in \Z+\a \\ \nu>0}} \log\left(1 - e^{2\pi i\b}e^{2\pi i \lambda \sum_{n=1}^N \a_{2n-1} (\nu\t)^{2n-1}}\right) - \delta_{\a,0}\delta_{\b,0}\frac{1}{2\lambda^2}\\
    =& \frac{\partial^2}{\partial \lambda^2} \int_0^\infty \log \left(1 - e^{2\pi i\b}e^{2\pi i\lambda \sum_{n=1}^N \a_{2n-1} (\t x)^{2n-1}}\right) dx \\
    &+ \frac{\partial^2}{\partial \lambda^2}\sum_{\substack{n\in\Z+\b\\ n>0}} \left(\sum_{\substack{z_j(n/\lambda) \\ \text{Im}(z_j(n/\lambda))>0}} \log\left(1-e^{2\pi i (\a + z_j(n/\lambda))}\right) + \sum_{\substack{z_j(n/\lambda) \\ \text{Im}(z_j(n/\lambda))<0}} \log\left(1-e^{2\pi i(\a - z_j(n/\lambda))}\right) \right)\;.
\end{split}\ee
We can now integrate \eqref{eq:psum} twice to obtain
\be\label{eq:sumwithAB}\begin{split}
    &\sum_{\substack{\nu \in \Z+\a \\ \nu>0}} \log\left(1 - e^{2\pi i\b}e^{2\pi i \lambda \sum_{n=1}^N \a_{2n-1} (\nu\t)^{2n-1}}\right) + \frac{1}{2}\delta_{\a,0}\delta_{\b,0}\log(\lambda) + A\lambda + B\\
    &=\int_0^\infty \log\left(1 - e^{2\pi i\b}e^{2\pi i\lambda \sum_{n=1}^N \a_{2n-1} (\t x)^{2n-1}}\right) dx \\
    &+ \sum_{\substack{n\in\Z+\b\\ n>0}} \left(\sum_{\substack{z_i(n/\lambda) \\ \text{Im}(z_i(n/\lambda))>0}} \log\left(1-e^{2\pi i (\a + z_i(n/\lambda))}\right) + \sum_{\substack{z_i(n/\lambda) \\ \text{Im}(z_i(n/\lambda))<0}} \log\left(1-e^{-2\pi i(\a + z_i(n/\lambda))}\right) \right)\;,
\end{split}\ee
where the two integration constants, $A$ and $B$, will depend on $\a,\b,\t$ and the $\a_{2n-1}$. We can fix these two constants by looking at the asymptotic behaviour of \eqref{eq:sumwithAB} as $\lambda \rightarrow \infty$. We will find that the asymptotic expansion will only have a term proportional to $\lambda$ and a constant term so we can then read off $A$ and $B$. (When $(\a,\b)=(0,0)$ we will also have a $\log(\lambda)$ term in the asymptotic expansion that cancels the one in \eqref{eq:sumwithAB}.)

We first note that the series
\be
    \sum_{\substack{\nu \in \Z+\a \\ \nu>0}} \log\left(1 - e^{2\pi i\b}e^{2\pi i \lambda  \sum_{n=1}^N \a_{2n-1} (\nu\t)^{2n-1}}\right)\;,
\ee
has a vanishing asymptotic expansion as $\lambda \rightarrow \infty$ since each term in the sum is exponentially suppressed. Hence it doesn't contribute to the constants $A$ and $B$.

The integral
\be
    \int_0^\infty \log\left(1 - e^{2\pi i\b}e^{2\pi i\lambda \sum_{n=1}^N \a_{2n-1} (\t x)^{2n-1}}\right) dx\;,
\ee
has an asymptotic expansion in powers of $1/\lambda$. To see this we first perform a change of variables in the integral. Recall that $z_0(x)$ is the root of \eqref{eq:zpoly} that vanishes when $x=0$. So if we change the integration variable to be $x = z_0\left(\frac{i u}{2\pi \lambda}\right)$ then the integral becomes
\be
    \int_0^\infty \left(\frac{d}{du}z_0\left(\tfrac{i u}{2\pi\lambda}\right)\right) \log(1 - e^{2\pi i\b}e^{-u}) du\;,
\ee
where we are able to move the integration contour back onto the positive real line since the only singularities in $u$ are on the imaginary axis which the contour doesn't cross. Integrating by parts then gives
\be
    - \int_0^\infty \frac{z_0\left(\tfrac{i u}{2\pi\lambda}\right)}{e^{2\pi i\b + u} - 1} du\;.
\ee
We can then expand $z_0(x)$ as a power series in $x$
\be
    z_0(x) = \sum_{n=1}^\infty z_{0,2n-1} x^{2n-1} \;,
\ee
and swap the sum and integral. Since the radius of convergence of the power series may be finite swaping the sum and integral leads to an expression that in general will be asymptotic. The asymptotic expansion of the integral as a power series in $\lambda$ is
\be
    \sum_{n=1}^\infty \frac{i(-1)^n z_{0,2n-1}}{(2\pi \lambda)^{2n-1}} \int_0^\infty \frac{u^{2n-1}}{e^{2\pi i\b + u}-1} du\;.
\ee
For $n \in \N$ and $\b = 0,\frac{1}{2}$ the integral can be evaluated in terms of the Hurwitz zeta function
\be\label{eq:zetaint}
    \int_0^\infty \frac{u^{2n-1}}{e^{2\pi i\b + u}-1} du = \frac{(-1)^n}{2}(2\pi)^{2n}\zeta(1-2n,1-\b)\;,
\ee
so we have the final asymptotic expansion for the integral in \eqref{eq:sumwithAB} as $\lambda \rightarrow \infty$
\be\label{eq:intasym}
    \int_0^\infty \log\left(1 - e^{2\pi i\b}e^{2\pi i\lambda \sum_{n=2}^N \a_{2n-1} (\t x)^{2n-1}}\right) dx \sim \pi i \sum_{n=1}^\infty \zeta(1-2n,1-\b) \frac{z_{0,2n-1}}{\lambda^{2n-1}} \;.
\ee
Finally we need to find the asymptotic expansion of 
\be
    \sum_{\substack{n\in\Z+\b\\ n>0}} \left(\sum_{\substack{z_j(n/\lambda) \\ \text{Im}(z_j(n/\lambda))>0}} \log\left(1-e^{2\pi i (\a + z_j(n/\lambda))}\right) + \sum_{\substack{z_j(n/\lambda) \\ \text{Im}(z_j(n/\lambda))<0}} \log\left(1-e^{-2\pi i(\a + z_j(n/\lambda))}\right) \right)\;,
\ee
as $\lambda \rightarrow \infty$. It will be easier to instead define $\tilde \lambda = 1/\lambda$ and find the asymptotic expansion as $\tilde\lambda \rightarrow 0^+$ of
\be\label{eq:sumtildelambda}
    \sum_{\substack{n\in\Z+\b\\ n>0}} \left(\sum_{\substack{z_j(n\tilde \lambda) \\ \text{Im}(z_j(n\tilde \lambda))>0}} \log\left(1-e^{2\pi i (\a + z_j(n\tilde \lambda))}\right) + \sum_{\substack{z_j(n\tilde \lambda) \\ \text{Im}(z_j(n\tilde \lambda))<0}} \log\left(1-e^{2\pi i(\a - z_j(n\tilde \lambda))}\right) \right)\;.
\ee
To do this we use the following results about asymptotic expansions from \cite{Zeidler2006} (which we reproduce in appendix \ref{sec:asym}). If we start with a function $f(t)$ which has the asymptotic expansion
\be
    f(t) \sim \sum_{n=0}^\infty b_n t^n \;,\quad t\rightarrow 0\;,
\ee
then we have the following asymptotic expansion
\be\label{eq:sumasym1}
    \sum_{m=0}^\infty f((m+a)t) \sim \frac{1}{t}\int_0^\infty f(x)dx + \sum_{n=0}^\infty \zeta(-n,a)b_n t^n \;,\quad t \rightarrow 0^+
\ee
where $\zeta(-n,a)$ is the Hurwitz zeta function \eqref{eq:Hurwitzzeta}. Since $\b=0,\frac{1}{2}$ we only need to consider the cases $a=\frac{1}{2},1$ in \eqref{eq:sumasym1} which leads to a simplification. Noting that
\be
    \zeta(-2n,a) = 0 \;,\quad n=1,2,\dots \;,\quad a=\frac{1}{2},1\;,
\ee
and
\be
    \zeta(0,1) = -\frac{1}{2} \;,\quad \zeta\left(0,\frac{1}{2}\right) = 0\;,
\ee
the asymptotic expansion \eqref{eq:sumasym1} becomes, for $a=\frac{1}{2},1$,
\be\label{eq:sumasym2}
    \sum_{m=0}^\infty f((m+a)t) \sim \frac{1}{t}\int_0^\infty f(x)dx - \frac{1}{2} b_0 \delta_{a,1} + \sum_{n=1}^\infty \zeta(1-2n,a)b_{2n-1} t^{2n-1} \;,\quad t \rightarrow 0^+\;.
\ee
The second result we need is for functions $f(t)$ which have a logarithmic term in their asymptotic expansions. If we have a function $f(t)$ with an asymptotic expansion of the form
\be\label{eq:logasym}
    f(t) \sim b\log(t) + \sum_{n=0}^\infty b_n t^n \;,\quad t\rightarrow 0\;,
\ee
then for $\a=\frac{1}{2},1$ as $t \rightarrow 0^+$ we have the asymptotic expansion
\be\label{eq:logasymsum}
    \sum_{m=0}^\infty f((m+a)t) \sim \frac{1}{t}\int_0^\infty f(x)dx + \frac{1}{2}\delta_{a,1}(b\log(2\pi/t)-b_0) + \frac{b}{2}\delta_{a,\frac{1}{2}}\log(2) + \sum_{n=1}^\infty \zeta(1-2n,a)b_{2n-1} t^{2n-1} \;.
\ee
We again prove this in appendix \ref{sec:asym}. This result can be generalised to a sum over the half lattice $m \in \N+a$ but we only need the special cases $a = 0,\frac{1}{2}$. The result for the general case is given in \cite{Zeidler2006}.

First we focus on the terms in \eqref{eq:sumtildelambda} that contain the $2N-2$ roots of \eqref{eq:zpoly} that don't vanish when $x=0$. As discussed in appendix \ref{sec:roots} these roots come in pairs which we label as $z_j^\pm(x)$, $j=1,\dots,N-1$ with Im$(z_j^+(x))>0$ and Im$(z_j^-(x))<0$ for $x>0$ and they are related via $z_j^+(x) = - z_j^-(-x)$. For a given pair $z_j^\pm(x)$ we define $f(t)$ to be
\be
    f(t) = \log\left(1 - e^{2\pi i(\a + z_j^+(t))}\right) + \log\left(1 - e^{2\pi i(\a - z_j^-(t))}\right)\;.
\ee
The function $f(t)$ is even and hence has the expansion in $t$
\be\label{eq:pmrootexpan}
    f(t) \sim 2\log\left(1 - e^{2\pi i(\a + z_j)}\right) + \sum_{n=1}^\infty b_{2n}t^{2n} \;,\quad t\rightarrow 0\;,
\ee
where $z_j^\pm(0) = \pm z_j$. Using the asymptotic expansion \eqref{eq:sumasym2} with \eqref{eq:pmrootexpan} gives
\be\label{eq:zisumasym}\begin{split}
    &\sum_{\substack{n\in\Z+\b\\ n>0}} \left(\log(1-e^{2\pi i (\a + z_j^+(n\tilde \lambda))}) + \log(1-e^{2\pi i(\a - z_j^-(n\tilde \lambda))}) \right)\\
    &\sim \frac{1}{\tilde\lambda} \int_0^\infty \left(\log(1 - e^{2\pi i(\a + z_j^+(x))}) + \log(1 - e^{2\pi i(\a - z_j^-(x))})\right) dx - \delta_{\b,0}\log(1 - e^{2\pi i(\a + z_j)})\;.
\end{split}\ee
The integral over $x>0$ can be written as two contour integrals
\be\label{eq:contourints}\begin{split}
    &\int_0^\infty \left(\log(1 - e^{2\pi i(\a + z_i^+(x))}) + \log(1 - e^{2\pi i(\a - z_i^-(x))})\right) dx \\
    &= \int_{z_j}^{z_j^+(\infty)} \log(1 - e^{2\pi i(\a + z)})\frac{dx}{dz}dz + \int_{-z_j}^{z_j^-(\infty)} \log(1 - e^{2\pi i(\a - z)})\frac{dx}{dz}dz\;,
\end{split}\ee
where the starting points of the contours are $z_j^\pm(0) = \pm z_j$. The end point of the contour tending to $z_j^+(\infty)$ is a ray tending to infinity along the direction $e^{i\theta^+}$, with $\theta^+ \in \left(\frac{\pi(2k-1)}{2N-1}, \frac{2\pi k}{2N-1}\right)$, $k=1,\dots,N-1$ and the end point tending to $z_j^-(\infty)$ is a ray long the direction $e^{i\theta^-}$, with $\theta^- \in \left(\frac{\pi(2k-1)}{2N-1}, \frac{2\pi k}{2N-1}\right)$, $k=N,\dots,2N-1$. To understand the end points of the contours we studied the asymptotic behaviour of the roots as $x \rightarrow \infty$ in appendix \ref{sec:roots}. Since $z_j^\pm(x)$ is a root of \eqref{eq:zpoly} the derivative $\frac{dx}{dz}$ is given by differentiating \eqref{eq:zpoly} with respect to $z$. Note that neither of the integration contours cross the real axis (again a result derived in appendix \ref{sec:roots}).

We integrate by parts and use the substitution $z=iw$ in the first integral and $z=-iw$ in the second integral
\be\label{eq:contourint1}
    -2\pi \int_{-i z_j}^{-i z_j^+(\infty)} \frac{x(iw)}{e^{2\pi (i\a + w)}-1}dw + 2\pi \int_{-i z_j}^{i z_j^-(\infty)} \frac{x(iw)}{e^{2\pi (i\a + w)}-1}dw\;.
\ee
In the second integral we have used $x(-iw) = -x(iw)$ as can be seen from \eqref{eq:zpoly}. These contours don't cross the imaginary axis where the integrand has poles. Since the contours both start at $-iz_i$ and end at $\infty$ along rays in the right hand side of the complex plane we can deform them to lie on top of each other and cancel. Hence the integral in \eqref{eq:zisumasym} vanishes. This then gives us the final asymptotic expansion
\be\label{eq:zisumasym1}
    \sum_{\substack{n\in\Z+\b\\ n>0}} \left(\log(1-e^{2\pi i (\a + z_j^+(n\tilde \lambda))}) + \log(1-e^{2\pi i(\a - z_j^-(n\tilde \lambda))}) \right) \sim - \delta_{\b,0}\log(1 - e^{2\pi i(\a - z_j)})\;.
\ee
We also have the root $z_0(x)$. This root vanishes at $x=0$, is an odd function and has negative imaginary part. So in \eqref{eq:sumtildelambda} we have the sum
\be
    \sum_{\substack{n\in\Z+\b\\ n>0}} \log(1-e^{2\pi i (\a - z_0(n\tilde \lambda))})\;.
\ee
Consider the function
\be\label{eq:z0f}
    f(t) = \log(1-e^{2\pi i(\a - z_0(t))}) \;.
\ee
The root $z_0(t)$ has a power series expansion
\be
    z_0(t) = t/\t + \sum_{n=2}^\infty z_{0,2n-1}t^{2n-1}
\ee
where we can determine the $t$ coefficient by substituting the expansion into the polynomial \eqref{eq:zpoly}. The functions
\be
    \log\left(\frac{1}{2}(1 + e^{-2\pi i z_0(t)})\right) + \pi i z_0(t) \;,\quad \log\left(\frac{\t}{2\pi i t}(1 - e^{-2\pi i z_0(t)})\right) + \pi i z_0(t)\;,
\ee
are both even and vanish at $t=0$. Hence the function $f(t)$ has the expansion
\be\label{eq:z0fasym}
    f(t) \sim \log(2) + \delta_{\a,0} \log(\pi i t/\t) - \pi i \sum_{n=1}^\infty z_{0,2n-1}t^{2n-1} + \sum_{n=1}^\infty a_{2n}t^{2n} \;.
\ee
When $\a=\frac{1}{2}$ we can use the asymptotic expansion \eqref{eq:sumasym2} which leads to
\be\label{eq:alphahalf}\begin{split}
    \sum_{\substack{n\in\Z+\b\\ n>0}} \log(1 + e^{-2\pi i z_0(n\tilde \lambda)}) \sim& \frac{1}{\tilde\lambda}\int_0^\infty \log(1 + e^{-2\pi i z_0(x)})dx - \frac{1}{2} \log(2) \delta_{\b,0} \\
    &- \pi i \sum_{n=1}^\infty \zeta(1-2n,1-\b) z_{0,2n-1} \tilde\lambda^{2n-1} \;,\quad \tilde\lambda \rightarrow 0^+\;.
\end{split}\ee
When $\a=0$ the asymptotic expansion of \eqref{eq:z0f}, given in \eqref{eq:z0fasym}, has a logarithmic term so we instead have to use the asymptotic result \eqref{eq:logasymsum}. This leads to
\be\label{eq:alpha0}\begin{split}
    \sum_{\substack{n\in\Z+\b\\ n>0}} \log(1 - e^{-2\pi i z_0(n\tilde \lambda)}) \sim& \frac{1}{\tilde\lambda}\int_0^\infty \log(1 - e^{-2\pi i z_0(x)})dx + \frac{1}{2}\delta_{\b,\frac{1}{2}}\log(2) - \frac{1}{2}\delta_{\b,0}\log(i\tilde\lambda/\t) \\
    &- \pi i \sum_{n=1}^\infty \zeta(1-2n,\frac{1}{2}) z_{0,2n-1} \tilde\lambda^{2n-1} \;,\quad \tilde\lambda \rightarrow 0^+\;.
\end{split}\ee
Hence combining \eqref{eq:alphahalf} and \eqref{eq:alpha0} we have as $\tilde\lambda \rightarrow 0^+$
\be\label{eq:z0sumasym}\begin{split}
    \sum_{\substack{n\in\Z+\b\\ n>0}} \log(1 - e^{2\pi i(\a - z_0(n\tilde \lambda))}) \sim& \frac{1}{\tilde\lambda}\int_0^\infty \log(1 - e^{2\pi i(\a - z_0(x))})dx + \frac{1}{2}(\delta_{\a,0}\delta_{\b,\frac{1}{2}} - \delta_{\a,\frac{1}{2}}\delta_{\b,0}) \log(2) \\
    &- \frac{1}{2}\delta_{\a,0}\delta_{\b,0}\log(i\tilde\lambda/\t) - \pi i \sum_{n=1}^\infty \zeta(1-2n,\b) z_{0,2n-1} \tilde\lambda^{2n-1} \;.
\end{split}\ee
We can explicitly evaluate the integral in \eqref{eq:z0sumasym}. We first write it as a contour integral like in \eqref{eq:contourints} and then use the same substitution to get an integral like the ones in \eqref{eq:contourint1}
\be
    \int_0^\infty \log(1 - e^{2\pi i(\a - z_0(x))})dx = 2\pi \int_0^{iz_0(\infty)} \frac{x(iw)}{e^{2\pi (i\a + w)}-1}dw\;.
\ee
The contour starts at the origin since $z_0(0)=0$ and tends to infinity in the right hand side of the complex plane. We can move the contour onto the positive real axis without picking up any poles since the contour never crosses the imaginary axis. Treating the polynomial \eqref{eq:zpoly} as a definition for the function $x(z)$ and the integral representation for the Hurwitz zeta function in \eqref{eq:zetaint} we can evaluate the integral and obtain
\be\begin{split}
    &\int_0^\infty \log(1 - e^{2\pi i(\a - z_0(x))})dx = 2\pi i\int_0^\infty \frac{\sum_{n=1}^N (-1)^{n+1}\a_{2n-1} (\t w)^{2n-1}}{e^{2\pi (i\a + w)}-1}dw \\
    &= -\pi i \sum_{n=1}^N \a_{2n-1} \t^{2n-1} \zeta(1-2n,1-\a) \;.
\end{split}\ee
Using this in the asymptotic expansion \eqref{eq:z0sumasym} gives
\be\label{eq:z0sumasym1}\begin{split}
    &\sum_{\substack{n\in\Z+\b\\ n>0}} \log(1 - e^{2\pi i(\a - z_0(n\tilde \lambda))}) \sim -\frac{\pi i}{\tilde\lambda} \sum_{n=2}^N \a_{2n-1} \t^{2n-1} \zeta(1-2n,1-\a)\\
    &+ \frac{1}{2}(\delta_{\a,0}\delta_{\b,\frac{1}{2}} - \delta_{\a,\frac{1}{2}}\delta_{\b,0}) \log(2) - \frac{1}{2}\delta_{\a,0}\delta_{\b,0}\log(i\tilde\lambda/\t) - \pi i \sum_{n=1}^\infty \zeta(1-2n,\b) z_{0,2n-1} \tilde\lambda^{2n-1} \;,
\end{split}\ee
as $\tilde\lambda \rightarrow 0^+$. Now if we combine \eqref{eq:intasym}, \eqref{eq:zisumasym1} and \eqref{eq:z0sumasym1} and express it all in terms of $\lambda = 1/\tilde\lambda$ we have the asymptotic expansion
\be\begin{split}
    &\int_0^\infty \log(1 - e^{2\pi i\b}e^{2\pi i\lambda \sum_{n=1}^N \a_{2n-1} (\t x)^{2n-1}}) dx \\
    &+ \sum_{\substack{n\in\Z+\b\\ n>0}} \left(\sum_{\substack{z_j(n/\lambda) \\ \text{Im}(z_j(n/\lambda))>0}} \log(1-e^{2\pi i (\a + z_j(n/\lambda))}) + \sum_{\substack{z_j(n/\lambda) \\ \text{Im}(z_j(n/\lambda))<0}} \log(1-e^{-2\pi i(\a + z_j(n/\lambda))}) \right)\\
    &\sim -\pi i \lambda \sum_{n=1}^N \a_{2n-1} \t^{2n-1} \zeta(1-2n,1-\a)  + \frac{1}{2}(\delta_{\a,0}\delta_{\b,\frac{1}{2}} - \delta_{\a,\frac{1}{2}}\delta_{\b,0}) \log(2)\\
    &+ \frac{1}{2}\delta_{\a,0}\delta_{\b,0}\log(-i\lambda\t) - \delta_{\b,0} \sum _{j=1}^{N-1} \log(1 - e^{2\pi i(\a + z_j)})\;,\quad \lambda \rightarrow \infty\;,
\end{split}\ee
where we recall that the $z_j$, $j=1,\dots,N-1$ are the non-vanishing roots to \eqref{eq:zpoly} with $x=0$ such that Im$(z_j)>0$. The higher order terms in $1/\lambda$ that appear in both \eqref{eq:intasym} and \eqref{eq:z0sumasym1} cancel so the final expansion only contains a term proportional to $\lambda$, a constant term and the necessary $\log(\lambda)$ term when $\a=\b=0$. Comparing this asymptotic expansion to \eqref{eq:sumwithAB} we can read off $A$ and $B$
\be\begin{split}
    &A = -\pi i \sum_{n=1}^N \a_{2n-1} \t^{2n-1} \zeta(1-2n,1-\a) \;,\\
    &B = \frac{1}{2}(\delta_{\a,0}\delta_{\b,\frac{1}{2}} - \delta_{\a,\frac{1}{2}}\delta_{\b,0}) \log(2) + \frac{1}{2}\delta_{\a,0}\delta_{\b,0}\log(-i\t) - \delta_{\b,0} \sum _{j=1}^{N-1} \log(1 - e^{2\pi i(\a + z_j)})\;.
\end{split}\ee
Using these values for $A$ and $B$ in \eqref{eq:sumwithAB} and setting $\lambda = 1$ gives
\be\label{eq:Htransform}\begin{split}
    &H_{\a,\b}(\t,\a_1,\dots,\a_{2N-1}) = \frac{1}{2}(\delta_{\a,\frac{1}{2}}\delta_{\b,0} - \delta_{\a,0}\delta_{\b,\frac{1}{2}}) \log(2) - \frac{1}{2}\delta_{\a,0}\delta_{\b,0}\log(-i\t)\\
    &+\int_0^\infty \log(1 - e^{2\pi i\b}e^{2\pi i \sum_{n=1}^N \a_{2n-1} (\t x)^{2n-1}}) dx + \delta_{\b,0} \sum _{j=1}^{N-1} \log(1 - e^{2\pi i(\a + z_j)})\\
    &+ \sum_{\substack{n\in\Z+\b\\ n>0}} \left(\sum_{\substack{z_j(n) \\ \text{Im}(z_j(n))>0}} \log(1-e^{2\pi i (\a + z_j(n))}) + \sum_{\substack{z_j(n) \\ \text{Im}(z_j(n))<0}} \log(1-e^{2\pi i(\a - z_j(n))}) \right)\;.
\end{split}\ee
where $H_{\a,\b}$ was defined in \eqref{eq:H} and as before the $z_i(x)$ are the roots to the polynomial \eqref{eq:zpoly} and the $z_i$ are the roots when $x=0$ such that Im$(z_i)>0$. Using $z_i^+(x) = -z_i^-(-x)$ we can rewrite the sum over the roots in the compact form
\be\begin{split}
    &\sum_{n\in\Z+\b} \sum_{\substack{z_j(n) \\ \text{Im}(z_j(n))>0}} \log(1-e^{2\pi i (\a + z_j(n))}) = \delta_{\b,0} \sum _{j=1}^{N-1} \log(1 - e^{2\pi i(\a + z_j)}) \\
    &+ \sum_{\substack{n\in\Z+\b\\ n>0}} \left(\sum_{\substack{z_j(n) \\ \text{Im}(z_j(n))>0}} \log(1-e^{2\pi i (\a + z_j(n))}) + \sum_{\substack{z_i(n) \\ \text{Im}(z_j(n))<0}} \log(1-e^{2\pi i(\a - z_j(n))}) \right)\;.
\end{split}\ee
If we then exponentiate the expression \eqref{eq:Htransform} we get the desired transformation of the generalised eta function, \eqref{eq:geneta},
\be\label{eq:transform11}\begin{split}
    &\eta_{\a,\b}(\t,\a_1,\dots,\a_{2N-1}) = \\
    &\sqrt{2}^{\delta_{\a,\frac{1}{2}}\delta_{\b,0} - \delta_{\a,0}\delta_{\b,\frac{1}{2}}} \sqrt{i/\t}^{\delta_{\a,0}\delta_{\b,0}}\exp\left( \int_0^\infty \log(1 - e^{2\pi i\b}e^{2\pi i \sum_{n=1}^N \a_{2n-1} (\t x)^{2n-1}}) dx \right)\\
    &\times \prod_{n\in\Z+\b} \prod_{\substack{z_j(n) \\ \text{Im}(z_j(n))>0}} (1-e^{2\pi i (\a + z_j(n))})\;,
\end{split}\ee
where the $z_j(x)$, $j=1,\dots,2N-1$ are the roots of the polynomial
\be\label{eq:poly1}
    \sum_{n=1}^N \a_{2n-1} (\t z)^{2n-1} = x\;.
\ee
This is exactly the transformation \eqref{eq:transform1} that we wanted to prove. Throughout the proof the variables $\t$ and $\a_{2n-1}$ were restricted to the domain given by Im$(\t)>0$, $\a_1\neq0$, Im$(\a_{2n-1}\t^{2n-1})\geq0$ and Im$(\a_{2N-1}\t^{2N-1})>0$. However both sides of \eqref{eq:transform11} are defined for the domain given by Im$(\t)>0$, $\a_1\neq0$ and Im$(\a_{2N-1}\t^{2N-1})>0$ so we can analytically continue the result into this larger domain. 

We can also consider the special case $\a_{2n-1} = 0$ for all $n<M\leq N$, $\a_{2M-1}\neq0$. We can find the transformation by taking \eqref{eq:transform11} with $\a_1\neq0$ and finding the limit $\a_1\rightarrow0$. The only terms that are affected by this limit are the $z_j(0)$ terms. If we naively set $\a_{2n-1} = 0$ for all $n<M\leq N$ and use the expression \eqref{eq:transform11} then the $2M-1$ roots of \eqref{eq:poly1} that vanish at $x=0$ will be lost. However if we solve the polynomial \eqref{eq:poly1} with $\a_1\neq0$ and take the limit $\a_1\rightarrow0$ then we have additional terms in \eqref{eq:transform11} corresponding to the roots that vanish at $x=0$.

\section{Infinite number of non zero $\a_{2n-1}$}\label{sec:infinite}

We end with a brief discussion on the case where an infinite number of the $\a_{2n-1}$ are non zero in the GGE \eqref{eq:GGE}. This corresponds to the case with an infinite number of $\a_{2n-1}$ variables in the generalised eta function \eqref{eq:geneta},
\be\label{eq:infiniteeta}\begin{split}
    \eta_{\a,\b}(\t,\a_1,\dots) =& \exp\left(-\pi i \sum_{n=1}^\infty \a_{2n-1} \t^{2n-1} \zeta(1-2n,1-\a)\right) \\
    &\times \prod_{\substack{k \in \Z+\a\\ k>0}}^\infty \left( 1 - e^{2\pi i (\b + \sum_{n=1}^\infty \a_{2n-1} (\t k)^{2n-1})} \right)\;.
\end{split}\ee
Since the sums in \eqref{eq:infiniteeta} are no longer finite we have to ensure that $\t$ and the $\a_{2n-1}$ are restricted to domains were they converge. In the product we have the series
\be
    \sum_{n=1}^\infty \a_{2n-1} (\t k)^{2n-1}\;.
\ee
Since the product is over $k \in \Z+\a, k>0$ we must restrict to domains of $\t$ and the $\a_{2n-1}$ where this series has an infinite radius of convergence. In \eqref{eq:infiniteeta} we also have the series
\be
    \sum_{n=1}^\infty \a_{2n-1} \t^{2n-1} \zeta(1-2n,1-\a)\;.
\ee
We again need to restrict to domains of $\t$ and the $\a_{2n-1}$ where this series converges. 

We define the function $F_{\t,\underline{\a}}(x)$, where $\underline{\a}$ is the set $\{\a_1,\a_3,\dots\}$, to be
\be
    F_{\t,\underline{\a}}(x) = \sum_{n=1}^\infty \a_{2n-1} (\t x)^{2n-1}\;.
\ee
Again we are only considering $\t$ and $\a_{2n-1}$ for which this series has an infinite radius of convergence. Since the radius of convergence of this series is infinite we can express the sum involving the zeta function as an integral using \eqref{eq:zetaint}
\be
    \sum_{n=1}^\infty \a_{2n-1} \t^{2n-1} \zeta(1-2n,1-\a) = \frac{1}{\pi i} \int_0^\infty \frac{F_{\t,\underline{\a}}\left(\frac{u}{2\pi i}\right)}{e^{2\pi i \a + u} - 1}du\;.
\ee
Hence the generalised eta function \eqref{eq:infiniteeta} can be expressed as 
\be\label{eq:etaF}
    \eta_{\a,\b}(\t,\a_1,\dots) = \exp\left(-\int_0^\infty \frac{F_{\t,\underline{\a}}\left(\frac{u}{2\pi i}\right)}{e^{2\pi i \a + u} - 1}du\right) \prod_{\substack{k \in \Z+\a\\ k>0}}^\infty \left( 1 - e^{2\pi i (\b + F_{\t,\underline{\a}}(k))} \right)\;.
\ee
Not only do we need to ensure that the function $F_{\t,\underline{\a}}(x)$ has a Taylor expansion with an infinite radius of convergence and the integral converges but we also need to ensure that the product converges i.e. $e^{2\pi i(\b + F_{\t,\underline{\a}}(k))}$ decays sufficiently rapidly as $k\rightarrow\infty$.

Assuming that the generalised eta function \eqref{eq:etaF} is well defined then we can again find a transformation formula for the generalised eta function under a modular transform $\t \rightarrow -1/\t$. We can follow the same steps as for the finite case. However before when computing the Fourier transform we had to find the roots of a polynomial, now we have to find the functions $z(x)$ that satisfy
\be
    F(z) = x\;.
\ee
This equation is equivalent to \eqref{eq:poly1} in the case with a finite number of non-vanishing $\a_{2n-1}$. The transformation will take the same schematic form as it does in \eqref{eq:transform11} but the product over the roots of the polynomial \eqref{eq:poly1} will be replaced with a product over the $z(x)$ which satisfy $F(z) = x$ and have Im$(z(x))>0$. 

Note that the factor $\sqrt{2}^{\delta_{\a,\frac{1}{2}}\delta_{\b,0} - \delta_{\a,0}\delta_{\b,\frac{1}{2}}}$ that appears in the transformation \eqref{eq:transform11} came from the root $z_0(x)$ of \eqref{eq:poly1} that vanishes when $x=0$. We were assuming throughout that $\a_1\neq0$ so we have only one such root. If the equation $F(z) = x$ has multiple solutions $z(x)$ that vanish when $x=0$ then this factor will be modified.

\section*{Acknowledgments}

I would like to thank S. Murthy and D. Zagier for discussions. I would like to thank G. Watts for discussions and reading early drafts of the manuscript. This work was supported by the EPSRC grant EP/V520019/1.

\appendix

\section{Roots}\label{sec:roots}
In this appendix we look at the roots of the polynomial
\be\label{eq:zxpoly}
    \sum_{n=1}^N \a_{2n-1} (\t z)^{2n-1} = x
\ee
as functions of $x$. When $x \in \Z+\b$, $\b=0,\frac{1}{2}$ these roots are the one particle energies in our transformed GGE. Here we show that these roots satisfy several properties that we make use of when proving the transformation of the GGE. The first is that the roots can be naturally paired up with one left over. When $x=0$ one of the roots vanishes and the others all take non zero values which come in pairs with opposite sign. This also holds for $x \neq 0$, there is one root which is an odd function of $x$ and vanishes at the origin, we label this root $z_0(x)$. The others come in pairs which we label $z_j^\pm(x)$, $j = 1,\dots,N-1$ and these pairs are related by $z_j^+(x) = -z_j^-(-x)$. Furthermore we show that for $x > 0$, Im$(z(x))\neq 0$ for all roots. So if Im$(z_j^+(0))>0$ then Im$(z_j^+(x))>0$ for all $x>0$ and hence Im$(z_j^-(x))<0$ for all $x>0$. We also show that Im$(z_0(x))<0$ for all $x>0$. Finally we find the asymptotic behaviour of the roots as $x\rightarrow\infty$ and see that they lie on the rays 
\be
    \frac{\exp(\frac{2\pi i k}{2N-1})}{(\a_{2N-1} \t^{2N-1})^{1/(2N-1)}}x^{\frac{1}{2N-1}}\;,\quad k=1,\dots,2N-1\;.
\ee
When $x=0$ one of the roots of \eqref{eq:zxpoly} is $0$ and the other roots are all non zero and come in pairs with opposite signs. ($\t \neq 0$ so we only have one vanishing root.) We label these non-vanishing roots $\pm z_j$, $j=1,\dots,N-1$ and fix Im$(z_j)>0$. When $x>0$ we label the roots by $z_0(x)$ and $z_j^\pm(x)$, $j=1,\dots,N-1$. Since the roots are continuous functions of $x$ we can connect them to the roots at $x=0$ as follows, at $x=0$ we have $z_0(0) = 0$ and $z_j^\pm(0) = \pm z_j$. 

Now we need to consider the roots when $x<0$. If $z(x)$ is a root of \eqref{eq:zxpoly} when $x<0$ then $-z(x)$ will be one of the roots $z_0(-x)$ or $z_j^\pm(-x)$. Hence we label the roots for $x<0$ the same as for $x>0$, $z_0(x)$ and $z_j^\pm(x)$, $j=1,\dots,N-1$. These roots are then defined by their relation to the roots for $x>0$. If $x<0$ then $z_0(x) = -z_0(-x)$ and $z_j^\pm(x) = -z_i^\mp(-x)$. This is then consistent with taking the limit $x\rightarrow0$ from either the positive or negative direction. Hence for $x \in \R$ we have the roots $z_0(x)$ and $z_j^\pm(x)$, $j=1,\dots,N-1$ which satisfy $z_0(x) = -z_0(-x)$ and $z_j^\pm(x) = -z_j^\mp(x)$. And when $x=0$ these roots are $z_0(0) = 0$ and $z_j^\pm(0) = \pm z_j$ where Im$(z_j)>0$.

We also need the result that if $z(x)$ is a root of \eqref{eq:zxpoly} then for $x>0$, Im$(z(x)) \neq 0$ (and by extension Im$(z(x)) \neq 0$ when $x<0$). We will prove this by contradiction. Assume that at $x=x_0>0$ we have Im$(z(x_0)) = 0$. If we plug this into the polynomial \eqref{eq:zxpoly} and take the imaginary part we have
\be\label{eq:Imofpoly}
    \sum_{n=1}^N \text{Im}(\a_{2n-1} \t^{2n-1}) z(x_0)^{2n-1} = 0\;.
\ee
However recall that throughout we are assuming that Im$(\a_{2n-1} \t^{2n-1}) \geq 0$, $n<N$ and Im$(\a_{2N-1} \t^{2N-1}) >0$ so \eqref{eq:Imofpoly} can only be satisfied if $z(x_0) = 0$ but then that implies $x_0 = 0$ which gives a contradiction.

Since Im$(z_i^+(0))>0$, as defined above, $z_i^+(x)$ will have positive imaginary part for all $x>0$ and similarly $z_i^-(x)$ will have negative imaginary part for all $x>0$. Finally we find that Im$(z_0(x))<0$ for $x>0$. The leading term in the expansion $z_0(x) = \frac{x}{\a_1\t} + \dots$ has negative imaginary part for $x>0$. Hence for small $x$, $z_0(x)$ will have negative imaginary part and since the roots don't cross the real axis it will have negative imaginary part for all $x>0$.

We also need the asymptotic value of the roots as $x \rightarrow \infty$. Assume that as $x \rightarrow \infty$ the roots $z(x)$ have leading order behaviour of the form $z \sim a x^\nu$, where $a$ is a constant and we must have $\nu>0$. Substituting this into \eqref{eq:zxpoly} gives the leading order behaviour
\be
    \a_{2N-1}(\t a x^\nu)^{2N-1} \sim x\;,
\ee
hence
\be
    \nu = \frac{1}{2N-1} \quad\text{and}\quad a = \frac{\exp(\frac{2\pi i k}{2N-1})}{(\a_{2N-1} \t^{2N-1})^{1/(2N-1)}} \;,\quad k = 1,\dots, 2N-1\;,
\ee
where since Im$(\a_{2N-1}\t^{2N-1})>0$ we choose the branch cut of $(\a_{2N-1} \t^{2N-1})^{1/(2N-1)}$ such that 
\be\label{eq:branch}
    \text{arg}((\a_{2N-1} \t^{2N-1})^{1/(2N-1)}) \in \left(0,\frac{\pi}{2N-1}\right)\;.
\ee
We have the leading asymptotic value of the roots
\be
    z_k^{(\infty)}(x) \sim \frac{\exp(\frac{2\pi i k}{2N-1})}{(\a_{2N-1} \t^{2N-1})^{1/(2N-1)}}x^{\frac{1}{2N-1}} \;,\quad x\rightarrow \infty \;,\quad k = 1,\dots, 2N-1\;,
\ee
where we choose the branch such that $x^{\frac{1}{2N-1}}$ is real for $x>0$. Note that with our branch choice \eqref{eq:branch} we have
\be
    \text{arg}(z_k^{(\infty)}(x)) \in \left(\frac{\pi(2k-1)}{2N-1},\frac{2\pi k}{2N-1}\right)\;,
\ee
hence
\be\begin{split}
    &\text{Im}(z_k^{(\infty)}(x)) > 0 \;,\quad k=1,\dots,N-1\;,\\
    &\text{Im}(z_k^{(\infty)}(x)) < 0 \;,\quad k=N,\dots,2N-1\;.
\end{split}\ee
So the asymptotic behaviour of the roots $z_j^+(x)$, $j=1,\dots,N-1$ will be of the form $z_k^{(\infty)}(x)$ for $k=1,\dots,N-1$ and the asymptotic behaviour of the roots $z_0(x)$ and $z_j^-(x)$, $j=1,\dots,N-1$, will be of the form $z_k^{(\infty)}(x)$ for $k=N,\dots,2N-1$.

\section{Asymptotic expansions}
\label{sec:asym}
In this appendix we reproduce some of the results from \cite{Zeidler2006} for the asymptotic expansions, as $t \rightarrow 0$ from above, of series of the form
\be\label{eq:fsum}
    \sum_{m=0}^\infty f((m+a)t)\;.
\ee
Let us start with a function $f(t)$ which has the asymptotic expansion
\be\label{eq:fasym}
    f(t) \sim \sum_{n=0}^\infty b_n t^n \;,\quad t\rightarrow 0\;.
\ee
We will further assume that $f(t)$ and all it's derivatives decay rapidly as $t\rightarrow \infty$. The asymptotic expansion of \eqref{eq:fsum} is
\be\label{eq:sumasym}
    \sum_{m=0}^{\infty}f((m+a)t) \sim \frac{1}{t}\int_0^\infty f(x)dx - \sum_{n=0}^\infty \frac{t^n B_{n+1}(a)}{(n+1)!} f^{(n)}(0) \;,\quad t \rightarrow 0^+\;,
\ee
where we are taking the limit $t \rightarrow 0$ from above. The Bernoulli polynomials, $B_n(x)$ are defined through their generating function in \eqref{eq:bernoullidef}. Alternatively using the relation \eqref{eq:bernoullizeta} between the Bernoulli polynomials and the Hurwitz zeta function \eqref{eq:Hurwitzzeta}, the asymptotic expansion is
\be\label{eq:sumasymzeta}
    \sum_{m=0}^\infty f((m+a)t) \sim \frac{1}{t}\int_0^\infty f(x)dx + \sum_{n=0}^\infty \zeta(-n,a)b_n t^n \;,\quad t \rightarrow 0^+\;,
\ee
where we have replaced the $f^{(n)}(0)$ with the $b_n$ coefficients from \eqref{eq:fasym}.

Our derivation will use the Bernoulli polynomials, $B_n(x)$, so we start with their generating function
\be\label{eq:bernoullidef}
    \frac{t e^{xt}}{e^t-1} = \sum_{n=0}^\infty \frac{B_n(x)}{n!}t^n\;.
\ee
We can use the generating function of the Bernoulli polynomials to prove the following identities that we will use when deriving \eqref{eq:sumasym},
\be\label{eq:bernoulliid}\begin{split}
    &B_n'(x) = n B_{n-1}(x) \\
    &B_n(x+1) - B_n(x) = nx^{n-1}\\
    &B_n(1-x) = (-1)^n B_n(x)
\end{split}\ee
The Hurwitz zeta function, which appears in \eqref{eq:sumasymzeta}, is defined by
\be\label{eq:Hurwitzzeta}
    \zeta(s,a) = \sum_{n=0}^\infty \frac{1}{(n+a)^s} \;,\quad \text{Re}(s)>0 \;,\; a \neq 0,-1,\dots
\ee
and can be analytically continued to $s \in\C\setminus{1}$. We have the relation between the Bernoulli polynomials and the Hurwitz zeta function, see \cite{Zeidler2006},
\be\label{eq:bernoullizeta}
    B_n(x) = -n\zeta(1-n,x)
\ee
which allows us to go between \eqref{eq:sumasym} and \eqref{eq:sumasymzeta}.

We begin the proof of \eqref{eq:sumasym} with the following integral identity. Using integration by parts and the identities \eqref{eq:bernoulliid} we have
\be\begin{split}
    &\int_m^{m+1} t^n f^{(n)}(xt)\frac{B_n(x-m+1-a)}{n!}dx \\
    =&  -\frac{(-t)^n B_{n+1}(a)}{(n+1)!}\left(f^{(n)}((m+1)t) - f^{(n)}(mt)\right) + \frac{f^{(n)}((m+1)t)}{n!}((1-a)t)^n\\
    &-\int_m^{m+1} t^{n+1}f^{(n+1)}(xt)\frac{B_{n+1}(x-m+1-a)}{(n+1)!}dx\;.
\end{split}\ee
Using induction on $n$ in the above identity we get
\be\label{eq:afternind}\begin{split}
    \int_m^{m+1} f(xt)dx =& \sum_{n=0}^{N-1} \frac{f^{(n)}((m+1)t)}{n!}((a-1)t)^n - \sum_{n=0}^{N-1} \frac{t^n B_{n+1}(a)}{(n+1)!}(f^{(n)}((m+1)t) - f^{(n)}(mt))\\
    & + \frac{(-t)^N}{N!} \int_m^{m+1} f^{(N)}(xt)B_N(x-m+1-a)dx\;.
\end{split}\ee
In order to replace the first sum with $f((m+a)t)$ we use Taylor's theorem. We can prove the following Taylor expansion with an integral remainder term by induction on $n$
\be
    \sum_{n=0}^{N-1} \frac{f^{(n)}((m+1)t)}{n!}((a-1)t)^n = f((m+a)t) + \frac{(-t)^{N-1}}{(N-1)!} \int_{(m+a)t}^{(m+1)t} (x/t-(m+a))^{N-1}f^{(N)}(x) dx\;.
\ee
Using this in \eqref{eq:afternind} gives
\be
    \int_m^{m+1} f(xt)dx = f((m+a)t) - \sum_{n=0}^{N-1} \frac{t^n B_{n+1}(a)}{(n+1)!}(f^{(n)}((m+1)t) - f^{(n)}(mt)) + t^{N-1}(r_1 + r_2)\;,
\ee
where the two remainder terms are
\be\begin{split}
    &r_1 = r_1(t,a,m) = \frac{(-1)^N t}{N!} \int_m^{m+1} f^{(N)}(xt)B_N(x-m+a)dx\;,\\
    &r_2 = r_2(t,a,m) = \frac{(-1)^{N-1}}{(N-1)!} \int_{(m+a)t}^{(m+1)t} (x/t-(m+a))^{N-1}f^{(N)}(x) dx\;.
\end{split}\ee
Now summing over $m=0,1,\dots,M-1$ gives
\be\label{eq:intermedstep}
    \int_0^M f(xt)dx = \sum_{m=0}^{M-1}f((m+a)t) - \sum_{n=0}^{N-1} \frac{t^n B_{n+1}(a)}{(n+1)!}(f^{(n)}(Mt) - f^{(n)}(0)) + t^{N-1}(R_1 + R_2)\;,
    \ee
where $R_i = R_i(t,a,M) = \sum_{m=0}^{M-1} r_i(t,a,m)$, $i=1,2$. 

We want to take the limit $M\rightarrow\infty$. We are assuming that $f(t)$ and all it's derivatives decay rapidly as $t\rightarrow\infty$, so all of the $f^{(n)}(Mt)$ terms will vanish. We also assume that the rapid decay of the derivatives is sufficient to ensure that the integral $\int_0^\infty |f^{(N)}(x)| dx$ converges for all $N$. We need to check that both of the remainder terms are finite in the limit $M\rightarrow\infty$ with $t$ fixed. 

We first look at $R_1$
\be
    R_1 = \frac{(-1)^N t}{N!} \sum_{m=0}^{M-1} \int_m^{m+1} f^{(N)}(xt)B_N(x-m+a)dx = \frac{(-1)^N }{N!} \int_0^{Mt} f^{(N)}(x)B_N(x/t-[x/t]+a)dx\;.
\ee
The function $|B_N(x/t-[x/t]+a)|$ is periodic and hence bounded, we denote the upper bound by $b_N$. Since we are also assuming $\int_0^\infty |f^{(N)}(x)| dx$ converges, in the limit $M\rightarrow\infty$ we can see that $R_1$ is finite
\be\label{eq:R1bound}
    \left|\lim_{M\rightarrow\infty} R_1\right| = \frac{1}{N!}\left| \int_0^\infty f^{(N)}(x)B_N(x/t-[x/t]+a)dx \right| \leq \frac{b_N}{N!} \int_0^\infty |f^{(N)}(x)|dx\;. 
\ee
Now we show that $R_2$ is bounded as $M\rightarrow\infty$. First we note that $|x/t-(m+a)|\leq |1-a|$ in the interval $x\in[(m+a)t,(m+1)t]$ so
\be
    |r_2(t,a,m)| \leq  \begin{cases}
        &\frac{|1-a|^{N-1}}{(N-1)!} \int_{(m+a)t}^{(m+1)t} |f^{(N)}(x)|dx \;,\quad a < 1\;,\\
        &\frac{|1-a|^{N-1}}{(N-1)!} \int_{(m+1)t}^{(m+a)t} |f^{(N)}(x)|dx \;,\quad a \geq 1\;.
    \end{cases}
\ee
When we sum over $m$ to obtain $R_2$ the regions of integration don't join up to give us an integral over the real line. We can however extend each integration domain to have two integers at the end points which will further bound $r_2$
\be
    |r_2(t,a,m)| \leq  \begin{cases}
        &\frac{|1-a|^{N-1}}{(N-1)!} \int_{(m+[a])t}^{(m+1)t} |f^{(N)}(x)|dx \;,\quad a < 1\;,\\
        &\frac{|1-a|^{N-1}}{(N-1)!} \int_{(m+1)t}^{(m+[a]+1)t} |f^{(N)}(x)|dx \;,\quad a \geq 1\;.
    \end{cases}
\ee
If we now sum over $m=0,1,2,\dots$ we can write these as integrals over the real line which will bound $R_2$ for us. We have add an extra factor to account for the overlap of the integral domains
\be\label{eq:R2bound}
    \left| \lim_{M\rightarrow\infty} R_2(t,a,M)\right| \leq \sum_{m=0}^\infty |r_2(t,a,m)| \leq \begin{cases}
        &\frac{|1-a|^{N-1}(|[a]|+1)}{(N-1)!}\int_{[a]t}^\infty |f^{(N)}(x)|dx \;,\quad a<1\;, \\
        &\frac{|1-a|^{N-1}[a]}{(N-1)!}\int_1^\infty |f^{(N)}(x)|dx \;,\quad a\geq1\;.
    \end{cases}
\ee
Hence upon taking the limit $M\rightarrow\infty$ and rearranging the terms in \eqref{eq:intermedstep} we have
\be
    \sum_{m=0}^\infty f((m+a)t) = \frac{1}{t}\int_0^\infty f(x)dx - \sum_{n=0}^{N-1} \frac{t^n B_{n+1}(a)}{(n+1)!}f^{(n)}(0) - t^{N-1} (\tilde R_1 + \tilde R_2)\;,
\ee
where $\tilde R_i = \tilde R_i(t,a) = \lim_{M\rightarrow\infty} R_i(t,a,M)$, $i=1,2$. If we now take the limit $t\rightarrow 0^+$ we can see from \eqref{eq:R1bound} and \eqref{eq:R2bound} that the two remainders $\tilde R_1(t,a)$ and $\tilde R_2(t,a)$ remain finite. We see that
\be
    \sum_{m=0}^\infty f((m+a)t) = \frac{1}{t}\int_0^\infty f(x)dx - \sum_{n=0}^{N-1} \frac{t^n B_{n+1}(a)}{(n+1)!}f^{(n)}(0) + O(t^{N-1})\;,
\ee
and hence we have the asymptotic expansion \eqref{eq:sumasym},
\be
    \sum_{m=0}^\infty f((m+a)t) \sim \frac{1}{t}\int_0^\infty f(x)dx + \sum_{n=0}^\infty \frac{(-t)^n B_{n+1}(a)}{(n+1)!}f^{(n)}(0) \;,\quad t \rightarrow 0^+\;.
\ee
Alternatively if we use identity \eqref{eq:bernoullizeta} relating the Bernoulli polynomials and the Hurwitz zeta function and replace the derivatives $f^{(n)}(0)$ with the $b_n$ via $n! b_n = f^{(n)}(0)$ we have the asymptotic expansion \eqref{eq:sumasymzeta}
\be
    \sum_{m=0}^\infty f((m+a)t) \sim \frac{1}{t}\int_0^\infty f(x)dx + \sum_{n=0}^\infty \zeta(-n,a)b_n t^n \;,\quad t \rightarrow 0^+\;.
\ee
We also need to consider the case where we have a function $f(t)$ with a logarithmic term in it's asymptotic expansion
\be\label{eq:flog}
    f(t) \sim b\log(t) + \sum_{n=0}^\infty b_n t^n \;,\quad t\rightarrow 0\;.
\ee
We only need to consider the special case where $a$ in \eqref{eq:fsum} is either $0$ or $\frac{1}{2}$. We will show that for $a=\frac{1}{2},1$ we have the asymptotic expansion
\be\label{eq:logsumasym}\begin{split}
    \sum_{m=0}^\infty f((m+a)t)\sim& \frac{1}{t}\int_0^\infty f(x)dx + \frac{1}{2}\delta_{a,1}(b \log(2\pi/t) - b_0) + \frac{b}{2}\delta_{a,\frac{1}{2}}\log(2) \\
    & + \sum_{n=1}^\infty \zeta(1-2n,a)b_{2n-1} t^{2n-1} \;,\quad t \rightarrow 0^+\;.
\end{split}\ee
To do this we instead look at the function
\be
    g(t) = f(t) - b \log(1 - e^{-2\pi t})\;.
\ee
The function
\be
    \log\left(\frac{1-e^{-2\pi t}}{2\pi t}\right) + \pi t\;,
\ee
is even and vanishes as $t\rightarrow 0$. Hence $\log(1 - e^{-2\pi t})$ has the asymptotic expansion
\be
    \log(1 - e^{-2\pi t}) \sim \log(2\pi t) - \pi t + \sum_{n=1}^\infty a_{2n}t^{2n}\;,
\ee
so this together with \eqref{eq:flog} gives the expansion for $g(t)$
\be
    g(t) \sim -b\log(2\pi) + \pi b t + \sum_{n=0}^\infty b_n t^n + \sum_{n=1}^\infty a_{2n} t^{2n} \;,\quad t \rightarrow 0^+\;.
\ee
Using \eqref{eq:sumasymzeta}, which we proved above, we have for $a=\frac{1}{2},1$
\be\label{eq:gsumasym}
    \sum_{m=0}^\infty g((m+a)t) \sim \frac{1}{t}\int_0^\infty g(x)dx + \frac{1}{2}\delta_{a,1}(b\log(2\pi) - b_0) + \pi \zeta(-1,a) bt + \sum_{n=1}^\infty \zeta\left(1-2n,\frac{1}{2}\right)b_{2n-1} t^{2n-1}\;,
\ee
as $t \rightarrow 0^+$, where we have used that for $a=\frac{1}{2},1$
\be
    \zeta(-2n,a) = 0\;,\quad n=1,2,\dots \;,\quad \zeta(0,a) = -\frac{1}{2}\delta_{a,1}\;.
\ee
The sums of $\log(1-e^{-2\pi t})$ can be written in terms of the theta, $\theta_i(\t)$, and Dedekind eta, $\eta(\t)$, functions which we define below. We can then make use of the known modular properties of these functions to get the asymptotic expansions of the sums of $\log(1-e^{-2\pi t})$. Subtracting this from the asymptotic expansion of $\sum_{m=0}^\infty g((m+a)t)$ gives the expansion of $\sum_{m=0}^\infty f((m+a)t)$. In terms of the $\theta_i(\t)$ and $\eta(\t)$ functions the sums of $\log(1-e^{-2\pi t})$ are 
\be\begin{split}
    &\sum_{m=0}^\infty \log(1 - e^{-2\pi (m+\frac{1}{2}) t}) = - \frac{\pi}{24}t + \frac{1}{2}\log \left( \frac{\theta_4(it)}{\eta(it)} \right)\;,\\
    &\sum_{m=1}^\infty \log(1 - e^{-2\pi m t}) = \frac{\pi}{12}t + \log (\eta(it))\;,
\end{split}\ee
where $\theta_4$ is the theta function
\be
    \theta_4(\t) = \prod_{n=1}^\infty \left(1-e^{2\pi i n \t}\right)\left(1-e^{2\pi i \left(n-\tfrac{1}{2}\right)\t}\right)^2\;,
\ee
and $\eta$ is the Dedekind eta function
\be
    \eta(\t) = e^{\frac{\pi i \t}{12}}\prod_{n=1}^\infty (1-e^{2\pi i n\t}) \;.
\ee
Using the modular properties of these two functions
\be\begin{split}
    &\eta(-1/\t) = \sqrt{-i\t} \eta(\t)\;,\\
    &\theta_4(-1/\t) = \sqrt{-i\t}\theta_2(\t)\;,
\end{split}\ee
where the branch cut is chosen such that $\sqrt{-i\t} = 1$ when $\t=i$ and $\theta_2$ is the theta function
\be
    \theta_2(\t) = 2e^{\frac{\pi i \t}{4}}\prod_{n=1}^\infty (1-e^{2\pi i n \t})(1+e^{2\pi i n\t})^2
\ee
we find
\be\begin{split}
    &\sum_{m=0}^\infty \log(1 - e^{-2\pi (m+\frac{1}{2}) t}) = -\frac{\pi}{24}t + \frac{1}{2}\log \left( \frac{\theta_2(i/t)}{\eta(i/t)} \right) \\
    &= -\frac{\pi}{24}t + \frac{1}{2}\log 2 - \frac{\pi}{12t} + \sum_{m=1}^\infty \log(1 + e^{-2\pi m/t}) \\
    &\sim -\frac{\pi}{24}t + \frac{1}{2}\log 2 - \frac{\pi}{12t} \;,\quad t \rightarrow 0^+ \;,
\end{split}\ee
and
\be\begin{split}
    &\sum_{m=1}^\infty \log(1 - e^{-2\pi m t}) = \frac{\pi}{12}t - \frac{1}{2}\log(t) + \log (\eta(i/t)) \\
    &= \frac{\pi}{12}t - \frac{1}{2}\log(t) - \frac{\pi}{12t} + \sum_{m=1}^\infty \log(1 + e^{-2\pi m/t}) \\
    &\sim \frac{\pi}{12}t - \frac{1}{2}\log(t) - \frac{\pi}{12t} \;,\quad t \rightarrow 0^+ \;,
\end{split}\ee
where we can drop the logarithm terms in the asymptotic expansion. Combining the two cases gives
\be\label{eq:logsumasym1}
    \sum_{m=0}^\infty \log(1 - e^{-2\pi (m+a) t}) \sim - \pi\zeta(-1,a)t + \frac{1}{2}(\delta_{a,\frac{1}{2}}\log(2) - \delta_{a,1}\log(t)) - \frac{\pi}{12t} \;,\quad t\rightarrow0^+\;.
\ee
We can also calculate the integral
\be
    \int_0^\infty \log(1 - e^{-2\pi x}) dx = - \frac{\pi}{12}\;,
\ee
which means
\be
    \int_0^\infty g(x) dx = \int_0^\infty f(x) dx + \frac{\pi}{12}b\;.
\ee
Adding the asymptotic expansions \eqref{eq:gsumasym} and \eqref{eq:logsumasym1} gives the asymptotic expansion of $\sum_{m=0}^\infty f((m+a)t)$,
\be\begin{split}
    &\sum_{m=0}^\infty f((m+a)t) = \sum_{m=0}^\infty g((m+a)t) + b\sum_{m=0}^\infty \log(1 - e^{-2\pi (m+a) t})\\
    &\sim \frac{1}{t}\int_0^\infty f(x)dx + \frac{1}{2}\delta_{a,1}(b \log(2\pi/t) - b_0) + \frac{b}{2}\delta_{a,\frac{1}{2}}\log(2) + \sum_{n=1}^\infty \zeta(1-2n,a)b_{2n-1} t^{2n-1} \;,\quad t \rightarrow 0^+\;,
\end{split}\ee
which is \eqref{eq:logsumasym}.

\bibliographystyle{JHEP}
\bibliography{Notes}
\end{document}